\documentstyle[12pt]{article}
\font\mygoth=eufm10 at 12pt
\newcommand\goth[1]{\hbox{\mygoth#1}}
\font\mybb=msbm10 at 12pt
\newcommand\bb[1]{\hbox{\mybb#1}}
\newcommand{\sect}[1]{\setcounter{equation}{0}\section{#1}}

\newcommand{\be}{\begin{equation}}
\newcommand{\en}{\end{equation}}
\newcommand{\bea}{\begin{eqnarray}}
\newcommand{\ena}{\end{eqnarray}}
\newcommand{\vs}[1]{\vspace{#1 mm}}
\newcommand{\nls}{non-linear Schr\"odinger }

\newcommand{\PL}{Phys. Lett. }
\newcommand{\NP}{Nucl. Phys. }
\newcommand{\W}{\mbox{\sf W}}

\newcommand{\Wi}{\mbox{${\sf W}_{1+\infty}$}}

\newcommand{\Q}{\mbox{$\tilde Q$}}
\newcommand{\h}{\mbox{$\tilde H$}}
\begin{document}
%
%
%
%
%
%
%
%
%
%
\renewcommand{\thefootnote}{\fnsymbol{footnote}}
\newpage
\setcounter{page}{0}
\pagestyle{empty}
\leftline{KCL-TH-92-2}
\leftline{July 1992}
\vs{20}

\begin{center}
{\LARGE {On the quantum KP hierarchy and its relation to the \nls
equation}}\\[1cm]
{\large M.D. Freeman and P. West}\\[0.5cm]
{\em King's
College, Department of Mathematics, Strand,
London WC2R 2LS, GB}
\\[1cm]
\end{center}
\vs{15}
\centerline{ \bf{Abstract}}
We establish a relation between the classical \nls equation and the KP
hierarchy, and we extend this relation to the quantum case by defining a
quantum KP hierarchy.  We also present evidence that an integrable
hierarchy of equations is obtained by quantizing the first Hamiltonian
structure of the KdV equation.  The connection between infinite-dimensional
algebras and integrable models is discussed.
\renewcommand{\thefootnote}{\arabic{footnote}}
\setcounter{footnote}{0}
\newpage
\pagestyle{plain}

%
%
%
\sect{Introduction}
Integrable dynamical systems with an infinite number of degrees
of freedom have been much studied for many years.  For such
dynamical systems, integrability is taken to mean the existence of an
infinite number of conserved quantities in involution, i.e. for which the
Poisson bracket between any pair is zero.
Examples of such sytems include the KdV equation,
\be
{\partial u \over \partial t} = {1\over 8} u^{\prime\prime\prime} -
{3\over 8} u u',
\en
the Liouville equation,
\be
\partial\bar\partial \phi - e^{-\phi} = 0,
\en
and the sinh-Gordon equation
\be
\partial\bar\partial \phi + 2 \sinh \phi = 0.
\en
These systems are in fact related to the Lie algebra $sl(2)$.
Corresponding to every Lie algebra there is an
integrable generalization of each of them \cite{DS,Wi},
these being the generalized KdV equations, Toda equations and affine Toda
equations respectively.

All of these equations can be written in Hamiltonian form. In each case the
Hamiltonian generating the time evolution has the property that it commutes
with all of the conserved charges with respect to the Poisson bracket.
Each equation is therefore a member of an infinite hierarchy of equations,
obtained by using the conserved charges as Hamiltonians to define commuting
time evolutions for infinitely many times.

In another development, which at first sight has no connection with
integrability, conformal field theories have been intensively studied
recently.  A theory that is Poincar\'e invariant possesses a conserved
energy-momentum tensor, and if such a theory is in addition conformally
invariant then the energy-momentum tensor is traceless.  In two dimensions
this implies that, in light-cone coordinates $(z,\bar z)$, the
component
$T \equiv T_{zz}$ is independent of $\bar z$,
\be
\bar\partial T = 0,
\en
and so any polynomial in $T$ is conserved.  Thus the integrals of such
polynomials are conserved charges for the theory.

It is a remarkable fact that all the integrable systems referred to above
can be linked to the infinite-dimensional conformal algebra and its
extensions.  Of course any conformally invariant theory in two dimensions
automatically possesses an infinite number of conserved quantities, as
argued above, but
in order to have integrability we require a set of conserved quantities
in involution.  Hence it is necessary to
select from the set of all polynomials in $T$ a subset for which the
Poisson bracket between any pair of charges is zero.  It turns out that
the charges
arrived at in this way are essentially the conserved charges of the
sinh-Gordon and KdV equations, even though these systems are not
conformally invariant.  A similar result holds for the generalizations of
these systems corresponding to arbitrary Lie algebras. For each Lie algebra
there is an extension of the conformal algebra, referred to generically as
a \W-algebra, and choosing a commuting subset of conserved quantities in
the \W-algebra corresponding to the Lie algebra \goth{g} gives the
conserved charges for the integrable systems corresponding to \goth{g}.

Let us spell out in more detail how this procedure works in the simplest
case.  The first step is to write the KdV equation in Hamiltonian form.  It
is well known that there are two possible Poisson brackets for the field
$u(x)$.  With respect to the so-called second Poisson bracket structure
\be
\{ u(x), u(y) \} = \left( u(x) \partial_x + \partial_x u(x) - \partial_x^3
\right)\delta(x-y)
\en
the time evolution is given by the Hamiltonian
\be
H_2 = {1\over 16} \int u^2,
\en
while in terms of the first Poisson bracket structure
\be
\{ u(x), u(y) \} = \partial_x \delta (x-y)
\en
the time evolution is given by
\be
H_3 = -{1\over 16} \int u^3 + u'^2.
\en
It is through the second Poisson bracket that the connection with the
conformal algebra is made.  It was recognized in \cite{G} that the second
Poisson bracket for the Fourier modes of $u$ is precisely the
two-dimensional conformal algebra, so that we can identify $u$ with the
energy-momentum tensor $T$.  The conserved quantities of the KdV equation
are then integrals of polynomials in $T$ and its derivatives.

If now we take a realization of the second Poisson bracket structure of the
KdV equation in terms of some field $\Phi$, so that $u$ is expressed in
terms of $\Phi$ and the Poisson bracket for $u$ follows from that of
$\Phi$, we can obtain an integrable hierarchy of evolution equations for
$\Phi$.  This hierarchy is obtained simply by taking the conserved charges
for the KdV equation, expressing them in terms of $\Phi$, and then using
these as Hamiltonians to generate time evolutions for $\Phi$.  Such a
hierarchy is obtained for every possible realization of the Virasoro
algebra.

An example of this is the  realization of the second Poisson bracket
structure of the KdV equation given by the Miura transformation.  The field
$u(x)$ is expressed in terms of a field $v(x)$ satisfying
\be
\{ v(x), v(y) \} = \partial_x \delta (x-y)
\en
through the transformation
\be
u={1\over2}v^2 + v'.
\en
The hierarchy of evolution equations obtained in this way for $v(x)$ is
precisely the mKdV hierarchy.  We can take this example one stage further
by taking $v$ to be the derivative of a scalar field $\phi$. In this case
the Poisson bracket of $u$ with $\int dx \exp(-\phi(x))$ is easily seen to
be
zero.  In fact the conserved quantities of the KdV equation, when expressed
in terms of $v$ through the Miura transformation, are even functions of
$v$, and so these charges must also commute with $\int dx \exp(\phi(x))$
and
hence with $\int dx \cosh \phi(x)$.  But $\int dx \cosh \phi(x) $ is the
Hamiltonian for
the sinh-Gordon equation, and so the conserved charges for the KdV equation
are also conserved for the sinh-Gordon equation.  Hence the KdV, mKdV and
sinh-Gordon equations have essentially the same conserved quantities
because their fields provide different realizations of the classical
Virasoro algebra.

There are analogues of this picture for other integrable systems.  In
particular let us consider the $N$'th KdV hierarchy, for variables
$u_i, i = 1,\ldots N,$ corresponding to the Lie algbera $sl(N+1)$.  The
equations of this hierarchy can be written in Hamiltonian form with Poisson
brackets that are a classical limit of the $\W_{(N+1)}$ algebra of Fateev
and Lukyanov \cite{FL}.  There is a realization of the $u_i$ in terms of
fields $\phi_j$ with Poisson brackets $\{\phi_i'(x), \phi_j'(y)\} =
\delta'(x-y)$ via the generalized Miura transformation
\be
\prod_i (\partial + h_i\cdot\phi) = \sum u_s \partial^{N-s},
\en
where the $h_i$ are the weights of the fundamental representation of
$sl(N+1)$.  The evolution equations for the $\phi_i$ generated by the
Hamiltonians of the $N$'th KdV hierarchy are those of the $N$'th mKdV
hierarchy, and these equations have the same conserved quantities as the
affine Toda field theory based on $sl(N+1)$.

The above considerations have all been classical, and it is natural to
consider whether the same picture holds after quantization.  The quantum
KdV, quantum mKdV and quantum sinh-Gordon equations were considered in
\cite{SY}, and the first few commuting conserved quantities were
constructed.  It was indeed found that for these quantities the connection
spelt out above held also at the quantum level. Thus the charges for the
quantum KdV equation, when expressed in terms of a scalar field through the
Feigin-Fuchs construction, were precisely the charges found for the quantum
mKdV and sinh-Gordon equations.  The first few conserved charges for these
systems
have also been found from the perspective of perturbations of conformal
field theories \cite{Z,EY,HM}.  The existence of conserved commuting
charges at all expected levels has argued in a free-field realization in
\cite{FF}, but the methods that are used classically to construct the
conserved charges and establish their properties have not so far been
carried over to the quantum theories.  The difficulties in doing this can
be traced to the necessity of normal-ordering the quantum conserved
currents and their products.

One approach to establishing the integrability of quantum theories would be
to attempt to exploit the remarkable connection spelt out above at the
classical level between infinite-dimensional algebras and integrability.
Indeed one might conjecture that this pattern is the general case, and that
every integrable system is associated with the Virasoro algebra, or an
extension of it, in this way.  Then different realizations of a given
algebra will lead to different integrable systems, but all such systems
originating from a given algebra will have the same conserved quantities.

Clearly, given an infinite set of commuting quantities, we can define an
integrable system by taking them as Hamiltonians.  In
this sense one can regard the infinite set of commuting quantities as the
primary object.  From this point of view the conjecture is that any such
set of quantities is associated with the Virasoro algebra or a \W-algebra.
One can also go further and conjecture that the commuting quantities can be
seen as a Cartan subalgebra of an even larger algebra.  Some evidence for
this point of view was given in \cite{FHW}, where a search for commuting
charges constructed from a $U(1)$ Kac-Moody current was carried out.  An
infinite number of sets of such commuting charges was seen to exist, but
it was found that, although the calculation was carried out entirely within
the framework of the $U(1)$ Kac-Moody algebra, each set of commuting
charges could be associated with an exceptional \W-algebra.

The motivation for this paper was to consider other integrable systems and
examine the extent to which they conformed with the above conjecture.  To
this end we examined the \nls equation.  We will show that there is a
relation between this equation and the KP hierarchy which is analagous to
that which exists between the mKdV and KdV hierarchies.  The essence of the
connection between the \nls equation and the KP hierarchy is that the first
Poisson bracket structure \cite{Wa} of the KP hierarchy is isomorphic
\cite{Ya,YW} to the algebra \Wi, which is a linear extension of the
Virasoro algebra containing a single spin-$i$ quasiprimary field for each
spin $i\ge 1$ \cite{PRS}, and this algebra has a realization in terms of a
complex scalar field $\psi,\psi^*$.

The quantum KP hierarchy is then defined using the quantum \Wi\ algebra,
and
it is shown to possess conserved quantities at least at the first few
levels.  The connection between the classical \nls equation and the KP
hierarchy spelt out above is shown to extend to the quantum analogues.

\sect{The relation between the KP hierarchy and the \nls equation}

The \nls equation is given by
\be
{\partial\psi\over\partial t} + \psi'' + 2 \kappa\, \psi^* \psi^2 = 0,
\en
where $\psi,\psi^*$ is a bosonic complex scalar field and $\kappa$ is a
coupling
constant.  This equation can be written in Hamiltonian form
$\dot\psi=\{\psi,H\}$ provided we adopt the Poisson brackets
\be
\{\psi^*(x),\psi(y)\} = \delta(x-y),
\quad\{\psi(x),\psi(y)\} = \{\psi^*(x),\psi^*(y)\}=0
\label{PBnls}
\en
and take $H=\int(\psi^*\psi''+ \kappa\,\psi^{*2}\psi^2)$.

It is useful to note that it is possible to assign weights to $\psi$ and
$\psi^*$ in a manner consistent with the Poisson bracket.  We take
$\partial$ and $\delta(x-y)$ to have weight 1, and  then eqn (\ref{PBnls})
implies that the product $\psi^*\psi$ must have weight 1.  If we now take
$\kappa$ to have weight 1, the Hamiltonian $H$ will be homogeneous of
weight 2.

The \nls equation possesses an infinite set of conserved quantities $Q_n$,
given by the formula
\be
Q_n = {1\over \kappa}\int dx\,\psi(x) Y_n(\psi,\psi^*),
\en
where
\be
Y_1 = \kappa\, \psi^*,\qquad
Y_{n+1}=Y_n'+\psi \sum_{k=1}^{n-1} Y_k Y_{n-k}\quad\hbox{for $n \ge 1$}.
\en
The first few such charges are
\bea
Q_1 &=& \int dx\,\psi^* \psi\nonumber\\
Q_2 &=& - \int dx\, \psi^* \psi'\nonumber\\
Q_3 &=& \int dx\,(\psi^* \psi'' + \kappa\, \psi^{*2} \psi^2);
\ena
it can be checked that these charges commute and so can be used to generate
commuting evolutions, giving a
hierarchy of equations.

The KP hierarchy consists of an infinite set of differential equations
which has, at first sight, no relation to the \nls equation.  This set of
equations is defined in terms of a pseudodifferential operator
$Q$ of the form
\be
Q = D + q_0 D^{-1} + q_1 D^{-2} + \ldots\quad.
\label{Q}
\en
Here $D$ denotes $\partial/\partial z$ and satisfies the following
generalization of the Leibniz rule:
\be
D^n f = \sum_{r=0}^\infty { n \choose r} f^{(r)} D^{n-r},\quad n \in
\bb{Z}.
\en
An infinite set of commuting time evolutions for the fields
$q_i$, $i=0,1,2,\ldots,$ is given by
\be
{\partial Q \over \partial t_p} = [ (Q^p)_+,Q], \quad p=0,1,2,\ldots
\label{KPevolution}
\en
where $(Q^p)_+$ is the part of $Q^p$ involving no negative powers of $D$.
These evolution equations can be written in Hamiltonian form.  There are
in fact a number of ways to do this, but for our purposes it is the
so-called first Hamiltonian structure that is needed.  The Poisson bracket
in this case is obtained from the method of coadjoint orbits applied to
the Lie algebra of differential operators.  For convenience we summarize
the essential elements of this method as applied to a general Lie algebra
$\goth{g}$.  The idea is to define Poisson brackets between functions on
the
dual space $\goth{g}^*$ of $\goth{g}$.  Every element $X$ of the Lie
algebra
$\goth{g}$
defines a natural linear function on $\goth{g}^*$ according to the formula
\be
X(\omega) \equiv \omega(X),\qquad \forall\quad \omega \in \goth{g}^*.
\en
Then, given $X,Y \in \goth{g}$, thought of as functions on $\goth{g}^*$, we
define their Poisson bracket to be the function on $\goth{g}^*$ given by
\be
\{X,Y\}(\omega) = \omega([X,Y]).
\en
Thus, on the left-hand side of this equation $X$ and $Y$ are considered as
functions on $\goth{g}^*$, while on the right-hand side they are thought of
as elements of the Lie algebra $\goth{g}$.

In the case at hand $\goth{g}$ is taken to be the Lie algbera of
differential operators, and the dual $\goth{g}^*$ can be identified with
the space of pseudodifferential operators of the form (\ref{Q}).  Given a
differential operator $X$, which we write for convenience in the form
$X=x_0(z) + D\, x_1(z) + \ldots + D^n\, x_n(z)$, where the derivative
$D\equiv{\partial\over\partial z}$ acts to the right, and a
pseudodifferential operator $Q$ as in eqn
(\ref{Q}), we take
\bea
Q(X) &\equiv& tr(Q X)\nonumber\\
&\equiv& \int dz\,\{\hbox{coefficient of $D^{-1}$ in $Q X$}\}\nonumber\\
&=& \int dz\,\sum_{i=0}^n q_i(z) x_i(z).
\ena
Taking $X(z) = D^i \delta(z-x)$ and
$Y(z) = D^j \delta(z-y)$,  we obtain
\bea
\{q_i(x),q_j(y)\} &\equiv& \{X,Y\}(Q)\nonumber\\
&=& Q([X,Y])\\
&=& \sum_{r=0}^j {j\choose r} \partial_x{}^r (q_{i+j-r}(x) \delta(x-y)) -
\sum_{r=0}^i {i \choose r} (-1)^r q_{i+j-r} \partial_x{}^r
\delta(x-y).\nonumber
\label{PB}
\ena
In evaluating the commutator $[X,Y]$ we made use of the relation
\be
f\, D^m = \sum_{r=0}^m (-1)^{m-r} {m \choose r} D^r f^{(m-r)}.
\en
The algebra (\ref{PB}) is the first Poisson bracket structure for the KP
hierarchy \cite{Wa}.  This algebra is in fact precisely \Wi\ with zero
central charge \cite{Ya,YW}; the $q_i$ are related to the usual basis for
\Wi, denoted by $V^i$ \cite{PRS,P}, by a linear
transformation.  It can be shown that the time
evolutions (\ref{KPevolution}) can be written as
\be
{\partial q_i \over \partial t_p} = \{q_i,H_{p+1}\}, \quad p=0,1,2,\ldots,
\en
where
\be
H_p = {1\over p} tr(Q^{p})
\en

Just as for the \nls equation, there is an assignment of weights to the
$q_i$ which is consistent with the Poisson bracket structure (\ref{PB}).
If we
take $\partial$ and $\delta(x-y)$ to have weight 1, then $q_i$ must have
weight $i+1$.  With this choice, the operator $Q$ in eqn (\ref{Q}) does not
have a definite weight, but we can correct this by working instead with
\be
\Q \equiv \kappa^{-1} D + q_0 D^{-1} + q_1 D^{-2} + \ldots,
\en
where, as before, $\kappa$ is a constant of weight 1.  \Q\ then has weight
zero.  Using
\Q\ instead of $Q$ does not change the Poisson brackets between the $q_i$,
but the Hamiltonians
\be
\h_p = {1\over p} tr(\Q^{p})
\en
now involve various powers of $\kappa$.  The first few $\h_p$ are given
explicitly by
\bea
\h_1 &=& \int dx\, q_0\nonumber\\
\h_2 &=& \kappa^{-1} \int dx\,q_1\\
\h_3 &=& \kappa^{-2} \int dx\,(q_2 + \kappa q_0{}^2)\nonumber\\
\h_4 &=& \kappa^{-4} \int dx\,(q_3 + 3 \kappa q_0 q_1)\nonumber.
\ena
The $\h_p$ can be viewed as a set of Poisson commuting quantities within
the enveloping algebra of \Wi.

Having given the KP and the \nls equations, we are now in a position to
explain the relation between them by exploiting the fact that \Wi\ has a
realization in terms of a single complex boson.  Let us take a general
approach first.  We consider a realization of \Wi\ in terms of some set of
fields $\phi_i$, so that the $q_i$ can be expressed in terms of the
$\phi_i$ and the Poisson brackets for the $q_i$ follow from those of the
$\phi_i$.  If such a realization of \Wi\ has zero central charge, we can
clearly construct an infinite set of mutually commuting quantities by
expressing the $H_n$'s in terms of the $\phi_i$.  We can then define time
evolutions for the $\phi_i$ by
\be
{\partial \phi_i \over \partial t_p} = \{\phi_i,H_{p+1}\}, \quad
p=0,1,2,\ldots,
\en
and it follows from the Leibniz property of the Poisson bracket that this
implies the KP evolutions for the $q_i$.

There is a centreless realization of \Wi\ in terms of a complex boson with
Poisson brackets given by
\be
\{\psi^*(x),\psi(y)\} = \delta(x-y),
\quad\{\psi(x),\psi(y)\} = \{\psi^*(x),\psi^*(y)\}=0.
\en
This is most easily seen by checking that the correct Poisson bracket
(\ref{PB})  between the $q_i$ is obtained by taking
\be
q_i = (-1)^i \psi^* \psi^{(i)}.
\en
The first few Hamiltonians of the KP hierarchy, when expressed in terms of
$\psi$ and $\psi^*$, take the forms
\bea
\h_1(\psi^*,\psi) &=& \int dx\,\psi^*\psi\nonumber\\
\h_2(\psi^*,\psi) &=& -  \kappa^{-1} \int dx\, \psi^*\psi'\\
\h_3(\psi^*,\psi) &=&  \kappa^{-2} \int dx\,
(\psi*\psi'' + \kappa\,\psi^{*2}\psi^2)\nonumber\\
\h_4(\psi^*,\psi) &=& -  \kappa^{-4} \int dx\,
(\psi^*\psi''' + 3 \kappa\, \psi^{*2}\psi\psi').\nonumber
\ena
We recognize these immediately as the first few conserved charges of the
\nls equation, up to overall factors.  It follows from the commutativity of
the KP Hamiltonians
that the $\h_n(\psi^*,\psi)$ will also commute.  However, since the charges
of the \nls equation are uniquely determined by the requirement that they
commute with $\h_3$, it follows that the charges obtained from the KP
system must coincide with those of the \nls equation at all levels.

\sect{A quantum KP hierarchy}

Given that the classical \nls equation is related to the KP hierarchy and
that the quantum \nls equation is known to be integrable, one might
expect there to exist an integrable quantum KP hierarchy.  One way to
construct such a
hierarchy would be
to take the first Poisson bracket structure for the KP hierarchy, which is
isomorphic to \Wi, and then to replace Poisson brackets by commutators.
The existence of an integrable quantum
KP hierarchy would then follow if we could find an infinite set of
commuting Hamiltonians in the enveloping algebra of \Wi.

In this section we construct some quantum analogues of the Hamiltonians
$\h_p$ from
the generators $V^i$ of \Wi.  We use the notation of \cite{PRS}, where
$V^i$ is a quasiprimary field of weight $i+2$.  The first step is to obtain
explicit expressions for the
operator product expansions for the $V^i$,
which are given in terms of generalized hypergeometric functions in
\cite{P}.  We used Mathematica \cite{W} to calculate these
hypergeometric
functions; the
first few operator product expansions are given by
\bea
V^{-1}(z) V^{-1}(w) &=& {c\over (z-w)^2} + \ldots\nonumber\\
V^{-1}(z) V^{0}(w) &=& {V^{-1}(w)\over (z-w)^2} + \ldots\nonumber\\
V^{0}(z) V^{0}(w) &=& {c/2 \over (z-w)^4} + {2 V^{0}(w)\over (z-w)^2} +
{V^0{}'(w)\over z-w} + \ldots\nonumber\\
V^{-1}(z) V^{1}(w) &=& {2V^{0}(w)\over (z-w)^2} + \ldots\nonumber\\
V^0(z) V^1(w) &=& {V^{-1}(w)\over (z-w)^4} + {3 V^1(w)\over (z-w)^2} +
{V^1{}'(w)\over z-w} + \ldots
\ena
We then looked for commuting charges that could be expressed as integrals
of sums of products of the fields in \Wi.  The
commutator of such charges is found in the usual way from the operator
product expansions; the integrals of fields $A$ and $B$ commute if the
single pole term in the operator product expansion of $A$ with $B$ is a
derivative.
We used the package OPEdefs of Thielemans \cite{T} to calculate the
necessary OPE's.  The first two charges, namely
\be
Q_1 = \int dz\, V^{-1}
\en
and
\be
Q_2 = \int dz\,V^0
\en
are trivial, in the sense that the first of these commutes with all
$V^i(z)$ while the commutator of $Q_2$ with $V^i(z)$ is just
$V^i{}'(z)$. For the first non-trivial charge
we take
\be
Q_3 \equiv \int dz\,\left\{V^1 + \kappa\,(V^{-1})^2\right\};
\en
this is analogous to the classical Hamiltonian $H_3$ of the KP hierarchy
in
that it involves terms of weights 2 and 3.  In this and subsequent
expressions involving products of operators, we use the normal ordering
standard in conformal field theory, namely we take
\be
(AB)(z) = \oint_z dw \, {A(w) B(z) \over w-z}.
\en
It is our belief that there are infinitely many further charges that
commute with the charge $Q_3$, and that furthermore these charges are
unique.
Using a
computer we have found three such charges, and we have verified that
all the charges we have obtained commute amongst themselves.  We
conjecture
that there exists an infinite set of such charges, one for each spin.  We
give here $Q_4$ and $Q_5$:
\bea
Q_4 &=& \int dz\,\left\{V^2 + 3 \kappa\,(V^{-1} V^0) + {3 c\,\kappa^2/2}\>
(V^{-1})^2\right\}\nonumber\\
Q_5 &=& \int dz\,\left\{ V^3 + \kappa\left( 4(V^{-1} V^1) + 2(V^0)^2 -
40/3\> (V^{-1}{}')^2\right)\right.\nonumber\\
&&\left. + \kappa^2 \left( 2(V^{-1})^3 + 4c\,(V^{-1}V^0)\right)
+ 2c^2\,\kappa^3\, (V^{-1})^2\right\}.
\ena

\sect{The quantum \nls equation}
The integrability of the quantum \nls equation has been extensively studied
using the quantum inverse scattering method and also from the point of view
of the existence of an infinite number of commuting conserved quantities.
The relation between these different approaches was discussed in
\cite{OSSY}.  In order to quantize the \nls equation we adopt the following
OPE for the bosonic operators $\psi$ and $\psi^*$:
\be
\psi^*(z) \psi(w) = (z-w)^{-1} + \ldots\quad.
\label{ope}
\en
The time evolution is generated by
\be
H = \oint dz \, (:\psi^*\psi'': + :\kappa \psi^{*2} \psi^2:),
\en
in the sense that
\be
{\partial\psi\over\partial \bar z} = [\psi,H],\quad
{\partial\psi^*\over\partial \bar z} = [\psi^*,H].
\en
We can then search for normal-ordered polynomials in $\psi,\psi^*$ and
their derivatives whose integrals commute with $H$; the first few conserved
currents are as follows:
\be
\psi^*\psi, \quad\psi^*\psi',\quad\psi^*\psi'' +
\kappa\psi^{*2}\psi^2,\quad\psi^*\psi'''+ 6\kappa\psi^*\psi''+
3\kappa\psi^{*2}\psi\psi'.
\en

In section 2 we showed that the \nls equation and the KP hierarchy were
related at the classical level, and we now extend that analysis to the
quantum case.  Just as the Poisson algebra \Wi\ has a realization in terms
of
classical scalar fields $\psi$ and $\psi^*$, the quantum algebra \Wi\ has a
realization in terms of operators $\psi$ and $\psi^*$ with OPE given by eqn
(\ref{ope}).  The quasiprimary fields $V^i$ of \Wi\ have the form
\be
V^i = {2n+2\choose n+1}^{-1}
\sum_{r=0}^{i+1} (-1)^r {i+1\choose r}^2 \psi^{(r)}\psi^{*(i+1-r)}.
\en
On substituting these expressions into the quantum KP charges $Q_i$ of the
previous section we recover the charges of the quantum \nls equation given
above, up to the freedom to add lower spin charges.  Some care must be
taken in transforming from the
normal ordering in terms of the $V^i$ to the free field normal ordering
used for the \nls charges.  Since the KP charges commute it follows that
the \nls charges will also commute, and because the \nls and KP charges are
uniquely determined by requiring that they commute with $H_3$ we conclude
that we obtain all the \nls charges in this way.

\sect{Quantizing the first Hamiltonian structure for the KdV equation}
We have seen that there is strong evidence that quantizing the first
Hamiltonian structure of the KP hierarchy leads to a quantum integrable
system.  This leads one to suspect that it may be possible to define an
alternative integrable quantum KdV system by quantizing the first Poisson
bracket structure $\{u(x),u(y)\}= \delta'(x-y)$ for this equation.  We
demand that $u$ be the current of a $U(1)$ Kac-Moody algebra and so satisfy
the OPE
\be
u(z) u(w) = (z-w)^{-2} + \ldots\quad.
\en
We consider a time evolution generated by $H=\oint (u'^2 + \kappa\,
u^3)$,
where the coupling constant $\kappa$ is included so that $H$ has definite
weight, and look for charges commuting with this operator.  We have found
three non-trivial charges commuting with $H$.  These charges also commute
with each other. The first two are
\bea
&&\oint dz\,\left\{ u''^2 + 5\, \kappa\, u u'^2 + 5\,\kappa^2/4\>
(u^4-u'^2)\right\}\nonumber\\
&&\oint dz\,\left\{u'''^2 + 7\, \kappa\, u u''^2 + 35\,\kappa^2/2\>
(u^2 u'^2 -1/6\,u''^2) + 7\,\kappa^3/4\>(u^5 - 5\,u u'^2)\right\};\\
\ena
the third is a polynomial of degree 7 in $\kappa$, with first term
$u''''^2$.
The charges found in this way
are not the same as those obtained by quantizing the second Poisson bracket
structure of the KdV equation.  It has already been noted \cite{KM} that
the first non-trivial equations of motion obtained from the first and
second Poisson bracket structures are not equivalent in the quantum case.
It is remarkable, however, that each appears to lead to an integrable
quantum hierarchy.

\sect{Conclusions}
We have shown that the classical \nls equation is obtained as a particular
realization of the KP hierarchy, and by quantizing the first Hamiltonian
structure of the KP hierarchy we have shown that the quantum \nls equation
can be obtained in a similar way.  It would be of great interest to
quantize the second Hamiltonian structure of the KP hierarchy.  The second
Poisson bracket structure is known \cite{2KP}, but it is nonlinear, and its
quantum analogue has not been found. Given this latter structure, however,
it would be straightforward to quantize the KP hierarchy by taking
commuting charges as Hamiltonians.  It would be interesting to study the
quantum Hamiltonian reduction of the quantum KP hierarchy to obtain the
quantum KdV equation; such a step would explain the structure of the
charges and possibly the integrability of the latter equation.  It would
also be interesting to investigate
further the relationship between the two quantum integrable systems
obtained by quantizing the two Hamiltonian structures for the KdV equation.

Acknowledgement: We are grateful to Klaus Hornfeck for many useful
discussions.  M.D. Freeman is grateful to the UK
Science and Engineering Research Council for financial support.

Note added: After this work was completed, Dr J. Gibbons informed us of a
connection between the \nls equation and the KdV equation.  We would like
to thank him for this information.
%
%

\end{document}